\documentclass[preprint]{aastex61}

\shorttitle{Kinematical properties of MPs in NGC 6362}
\shortauthors{Dalessandro et al.}

\begin{document}

\title{The unexpected kinematics of multiple populations in NGC~6362: do binaries play a role?\footnote{Based on data obtained with the Very Large Telescope 
at the European Southern Observatory, programs: 093.D-0618 and 097.D-0325 (PI: Dalessandro)}}

\correspondingauthor{Emanuele Dalessandro}
\email{emanuele.dalessandro@oabo.inaf.it}


\author{E. Dalessandro}
\affiliation{INAF-Astrophysics and Space Science Observatory, Via Gobetti 93/3 40129 Bologna - Italy}

\author{A. Mucciarelli}
\affiliation{Dipartimento di Fisica \& Astronomia, Universit\'a degli Studi di Bologna, Via Gobetti 93/2 40129 Bologna -
Italy}
\affiliation{INAF-Astrophysics and Space Science Observatory, Via Gobetti 93/3 40129 Bologna - Italy}

\author{M. Bellazzini}
\affiliation{INAF-Astrophysics and Space Science Observatory, Via Gobetti 93/3 40129 Bologna - Italy}

\author{A. Sollima}
\affiliation{INAF-Astrophysics and Space Science Observatory, Via Gobetti 93/3 40129 Bologna - Italy}

\author{E. Vesperini}
\affiliation{Department of Astronomy, Indiana University, Swain West, 727 E. 3rd Street, IN 47405 Bloomington - USA}

\author{J. Hong}
\affiliation{Department of Astronomy, Indiana University, Swain West, 727 E. 3rd Street, IN 47405 Bloomington - USA}
\affiliation{Kavli Institute for Astronomy and Astrophysics, Peking University, Yi He Yuan Lu 5, HaiDian District, Beijing 100871, China}

\author{V. H\'enault-Brunet}
\affiliation{National Research Council, Herzberg Astronomy \& Astrophysics, 5071 West Saanich Road, Victoria, BC, V9E 2E7, Canada}

\author{F.R. Ferraro}
\affiliation{Dipartimento di Fisica \& Astronomia, Universit\'a degli Studi di Bologna, Via Gobetti 93/2 40129 Bologna -
Italy}
\affiliation{INAF-Astrophysics and Space Science Observatory, Via Gobetti 93/3 40129 Bologna - Italy}

\author{R. Ibata}
\affiliation{Observatoire Astronomique, Universit\'e de Strasbourg, CNRS, 11 rue de l'Universit\'e, F-67000 Strasbourg, France}

\author{B. Lanzoni}
\affiliation{Dipartimento di Fisica \& Astronomia, Universit\'a degli Studi di Bologna, Via Gobetti 93/2 40129 Bologna -
Italy}
\affiliation{INAF-Astrophysics and Space Science Observatory, Via Gobetti 93/3 40129 Bologna - Italy}

\author{D. Massari}
\affiliation{Kapteyn Astronomical Institute, University of Groningen, Groningen, The Netherlands
Leiden Observatory, Leiden University, Leiden, The Netherlands}

\author{M. Salaris}
\affiliation{Astrophysics Research Institute, Liverpool John Moores University, IC2 Liverpool Sceince Park, 146 Brownlow Hill, L3 5RF, Liverpool}




\begin{abstract}
We present a detailed analysis of the kinematic properties of the multiple populations (MPs)
in the low-mass Galactic globular cluster NGC~6362 based
on a sample of about 500 member stars
for which radial velocities (RVs), Fe and Na abundances have been homogeneously derived.

At distances from the cluster center larger than about $0.5 r_h$, we find that first (FG - Na-poor) 
and second generation (SG - Na-rich) stars show hints of different
line-of-sight velocity dispersion profiles, with FG stars being dynamically hotter.
This is the first time that differences in the velocity dispersion of MPs are detected by using only RVs.
While kinematic differences between MPs in globular clusters 
are usually described in terms of anisotropy differences driven by the different radial distributions, 
this explanation seems hardly viable for NGC~6362, where SG and FG stars are spatially mixed. We demonstrate that the observed difference in the velocity dispersion profiles can be accounted for by the effect of binary stars. In fact, thanks to our multi-epoch RV measurements we find
that the binary fraction is significantly larger in the FG sample ($f\sim14\%$) than in SG population ($f<1\%$), and we show that such a difference can inflate the velocity dispersion of FG with respect to SG by the observed amount in the relevant radial range.
Our results nicely match the predictions of state-of-the art $N$-body simulations of the co-evolution of MPs in globular clusters that include the effect of binaries.

\end{abstract}




\section{Introduction} \label{sec:intro}

The discovery of multiple populations (MPs) in globular clusters (GCs),
differing in light-elements abundance (e.g. 
He, C, N, O, Na, Mg, Al) while having the same iron (and iron-peak elements) content, 
has seriously challenged our understanding 
of the physical mechanisms driving the formation and early evolution of these systems (see \citealt{gratton12,bastian_lardo17} for a review).
Indeed it is now well established that almost all relatively massive ($\sim10^4 M_{\odot}$; e.g. \citealt{piotto15,bragaglia17}) 
and old ($> 2$ Gyr; see for example \citealt{martocchia17}) GCs host MPs.

Spectroscopically MPs manifest themselves in the form of light-element anti-correlations (like C-N, Na-O, Mg-Al). 
These chemical inhomogeneities produce also a variety of features in the color-magnitude diagrams (CMDs) 
when appropriate
near-UV bands are used \citep{sbordone11}.
Thanks to spectro-photometric studies, MPs have been directly observed in many GCs
in the Galaxy (see \citealt{piotto15} for a recent homogeneous collection) as well as in external systems (like the Magellanic Clouds and the Fornax dwarf galaxy - 
\citealt{mucciarelli09,dalessandro16,larsen14}). 
Moreover, the presence of MPs has been indirectly constrained in the GC systems of M31 and M87 \citep{schiavon13,chung11}. 

Different scenarios have been proposed over the years to explain the formation of MPs. They generally invoke 
a self-enrichment process, which likely occurred in the very early epochs of GC formation and evolution. 
In these scenarios, it is thought that a second generation/population (SG) formed from the ejecta of stars (polluters) 
to a first generation/population (FG) 
mixed with  ``pristine material'' \citep{decressin07,dercole08,demink09,bastian13,denis14}. 
However, all models proposed so far face serious problems 
and a self-consistent explanation of the physical processes at the basis of MP formation is still lacking.

Understanding the kinematical properties of MPs can provide new important insights on GC formation and evolution.
One of the predictions of MP formation models (e.g. \citealt{dercole10}) is that SG stars form a low-mass 
and centrally segregated stellar sub-system  possibly characterized by a more rapid internal rotation than the (more spatially extended) FG system \citep{bekki10}.
Although the long-term dynamical evolution of stars can smooth out the initial structural and
kinematical differences between FG and SG to a large extent, some are expected 
to be still visible in present-day GCs (see for example \citealt{vesperini13,VHB15}).
First evidence of the difference in the structural properties of MPs were based on their spatial distributions \citep{lardo11,milone12,bellini13,
dalessandro16,massari16}. These works 
have shown that indeed SGs are typically more centrally concentrated than FG sub-populations, with few remarkable exceptions
(see \citealt{dalessandro14,larsen15,savino18}). 
However, spatial distributions alone provide only a partial picture of the dynamical properties of MPs and key constraints on the possible formation and dynamical paths of MPs may be hidden in their kinematic properties.

Hints of different degrees of orbital anisotropy among MPs
have been found in the massive GCs 47 Tuc and NGC 2808 by means of HST proper motions \citep{richer13,bellini15}. These findings seem to
be consistent with the kinematical fingerprints of the diffusion of the SG population from the innermost
regions towards the cluster outer zones and provide indirect support to formation scenarios predicting SG formed in a centrally concentrated subsystem.

\citet{CorderoVHB2017} found also that the extremely enriched component of SG stars 
in M~13 shows a larger rotation amplitude than other stars in the cluster.

In general, however, MP kinematics is still poorly constrained, mainly because of technical limitations related to the difficulty of
deriving kinematical information in dense environments for large and significant samples of resolved stars adequately separated in terms of their light elements abundances.

Here we present the results of an extended kinematical analysis of the MPs in the Galactic GC NGC~6362,
based on the radial velocities (RVs) of a large sample ($\sim800$) of stars chemically tagged according to the sub-population they belong to.
The case of NGC~6362 is particularly interesting. Indeed, this is possibly the second least massive ($M\sim5\times10^4 M_{\odot}$) GC\footnote{after NGC~6535 which has $M\sim2\times10^4 M_{\odot}$;  
\citealt{piotto15,bragaglia17}} where 
MPs have been identified both photometrically and spectroscopically \citep{dalessandro14,mucciarelli16} so far. Also, 
contrary to what is generally observed in other GCs 
(see, e.g., \citealt{lardo11}), we have found that in this cluster the spatial distributions of MPs are consistent with complete mixing  over the entire cluster extension \citep{dalessandro14}.
This  behavior suggests that the cluster underwent 
 complete spatial redistribution of stars and severe mass-loss due to long-term dynamical evolution \citep{vesperini13,dalessandro14,miholics15}.  

The paper is structured as follows. In Section~2, the data-base and data-reduction are presented, in Section ~3 
sample selection criteria are described, while in Section~4 we report on the main results 
of the kinematic analysis. In Section~5 sub-population binary fractions are derived and their impact is estimated by means of an analytic approach. In Section~6 observations are compared to $N$-body models following the evolution of MPs. Conclusions and discussion are presented in Section~7.

\begin{figure}
\centering
\plotone{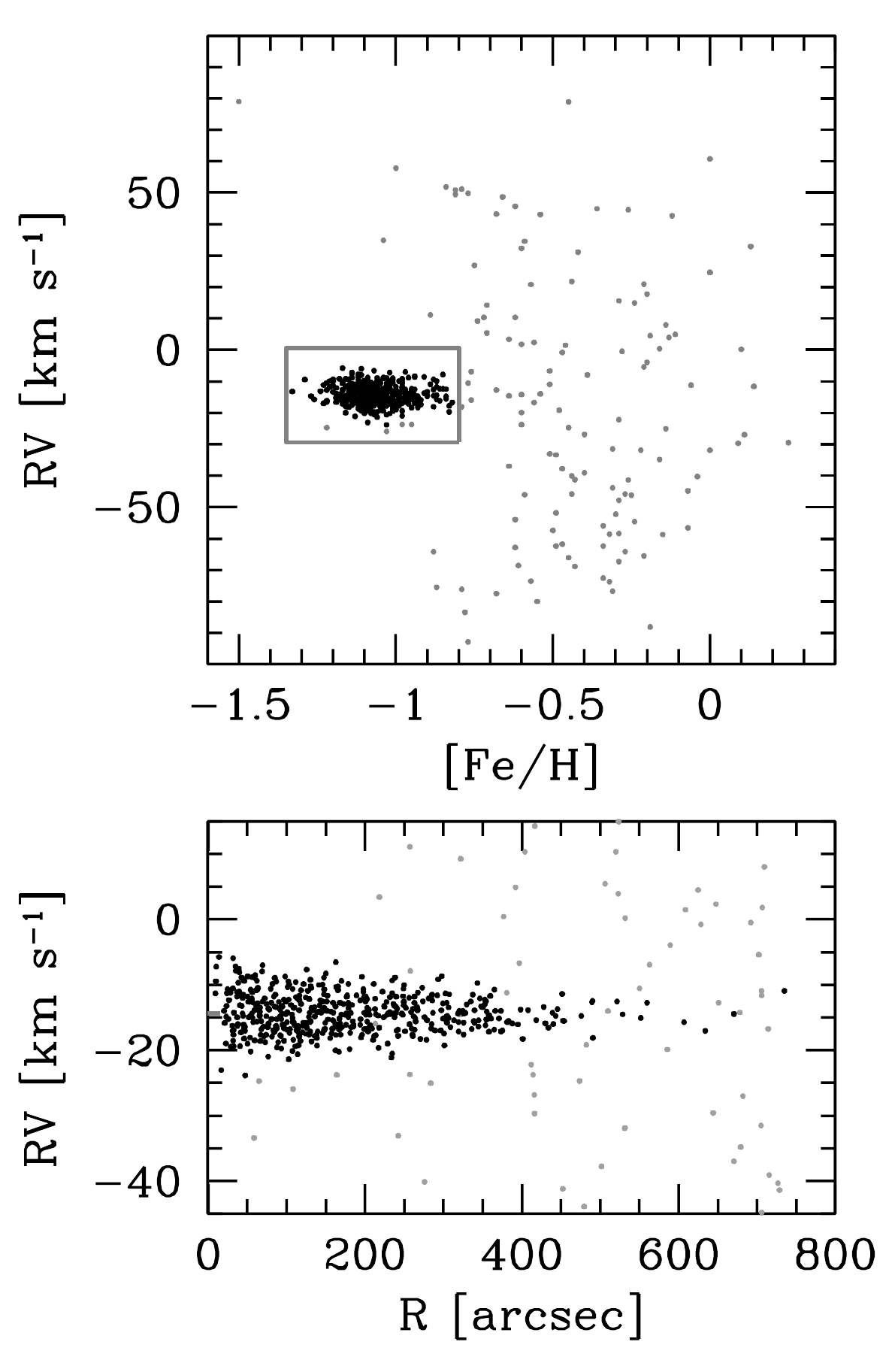}
\caption{Distribution the RVs of the stars in our sample as a function of the derived [Fe/H] (upper panel) and projected distance from 
the cluster center (bottom panel). Black dots represent stars selected as described in Section~3 and used for the kinematic analysis.}
\end{figure}

\section{Observations and data-analysis}
\subsection{Radial velocities and chemical abundances}

The dataset used in this work consists of spectra obtained 
using the multi-object facility FLAMES@ESO-VLT \citep{pasquini00} in the 
UVES+GIRAFFE combined mode and secured in two observing runs. 
In the first run (Prop. ID: 093.D-0618, PI: Dalessandro) stars have been 
observed using the GIRAFFE setups HR11 and HR13, sampling 
the two Na doublets at 5682-5688 $\AA$ and 6154-6160 $\AA$. In the second run (Prop. ID: 097.D-0325, PI: Dalessandro), 
which was devoted to enlarge the sample of available RVs and [Na/Fe] abundances toward fainter 
magnitudes, we adopted the GIRAFFE setup HR12, in order 
to cover the Na doublet at 5895-5890 \AA\ that is strong enough 
to be easily measured also at low Signal-to-Noise ratio (SNR).
All the UVES targets of both runs have been observed with 
the UVES Red Arm CD\#3 580 set-up.
Only red giant branch (RGB) stars brighter than $V\sim17.6$  and 
red horizontal branch (RHB) stars have been selected by using the optical and ultraviolet ground-based photometry 
by \citet{dalessandro14}.
219 stars were observed with the first run and 585 with the second one, 84 of them are in common between
the two runs, thus yielding a total of 720 observed stars in the two observing programs.
Many stars in the final sample have been observed up to six times.
All the spectra have been reduced with the dedicated ESO pipelines.
Some results obtained with the first run have been presented in \citet{mucciarelli16} and \citet{massari17}.

RVs have been measured for each individual spectrum 
with the code {\tt DAOSPEC} \citep{stetson08} using tens of metallic lines. 
Heliocentric corrections have been applied to each exposure, then the spectra of each target have been co-added 
together and used for the chemical analysis. In the analysis presented below we adopt for each star
the RV estimate obtained from the spectrum with the highest SNR. 
This allows us to incorporate homogeneously into the final sample the large number of stars having just one spectrum  
(156) with those having two or more spectra. 

Comparing the distribution of the standard deviations ($\sigma_{RV}$) as a function of magnitude, 
for the stars having at least four independent $RV$ estimates  
with the distribution of errors on individual best-SNR measures ($\epsilon_{RV}$, as derived by {\tt DAOSPEC}), 
we noted that, at any given magnitude, $\epsilon_{RV}$ was systematically smaller than $\sigma_{RV}$ (see 
\citealt{kirby15}). 
We found that multiplying $\epsilon_{RV}$ by 2 lead to a nearly perfect coincidence of the two distributions, 
hence we decided to apply this rescaling to get more reliable values of $\epsilon_{RV}$. 
These rescaled velocity errors have typical values of 
$\simeq  0.3$~km s$^{-1}$ for $V\le 15.0$, $\simeq  0.8$~km s$^{-1}$ for $15.0<V\le 17.0$, and $\simeq  1.2$~km s$^{-1}$ for $V> 17.0$.

Stellar atmospheric parameters have been determined as in \citet{mucciarelli16}. Briefly,
effective temperatures and surface gravities have been derived from the available photometry;
microturbolent velocities have been obtained spectroscopically. 

\begin{figure}
\plotone{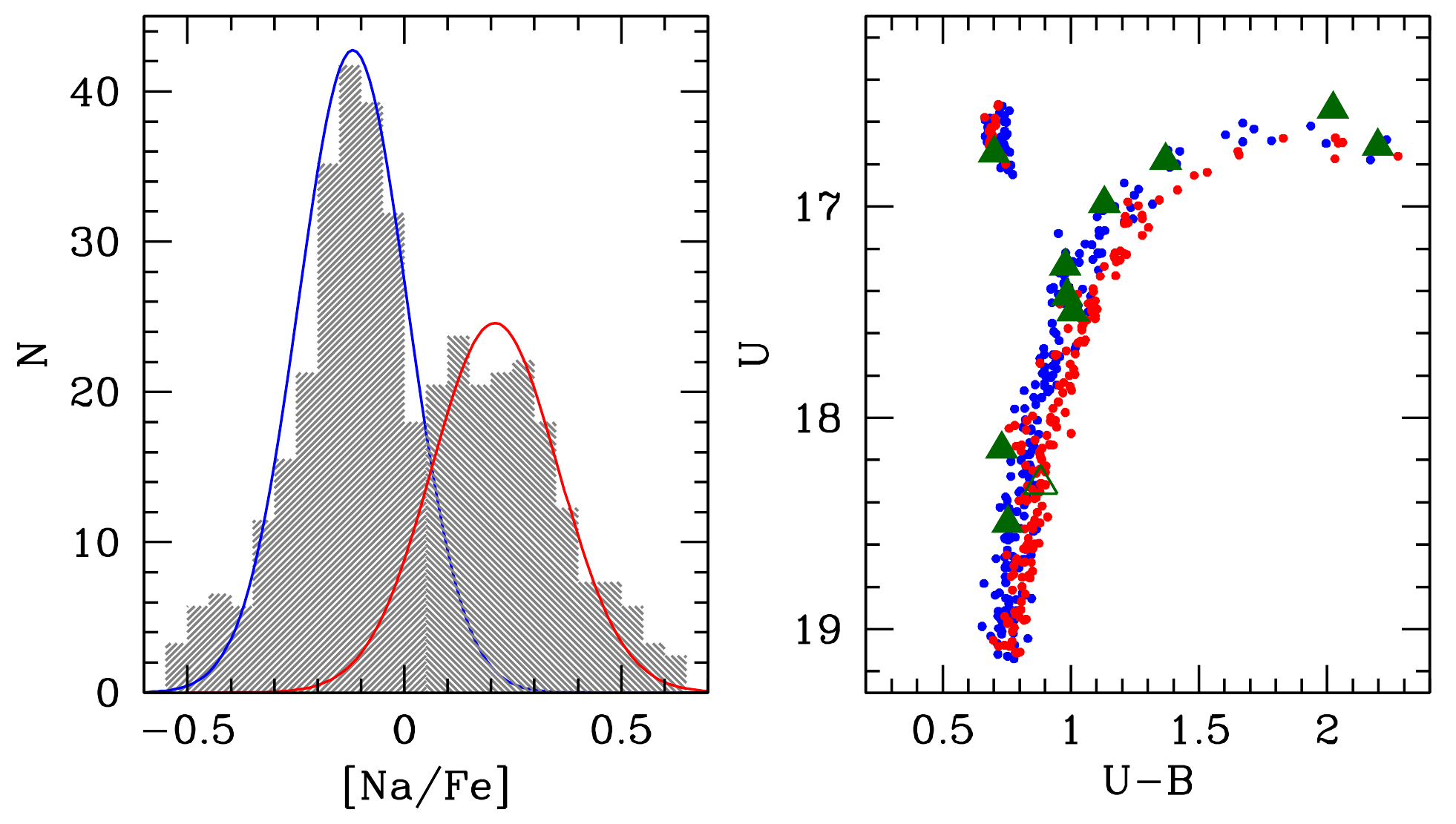}
\caption{{\it Left panel}: Distribution of the [Na/Fe] abundances of the 489 selected stars. 
{\it Right panel}: (U,U-B) CMD of the selected samples (photometry from Dalessandro et al. 2014); 
{\it Na-rich} stars are in red and {\it Na-poor} ones are in blue. Filled green triangles are 
{\it Na-poor} binary stars directly detected as described in Section~5, the open triangle is a {\it Na-rich}
binary.}
\end{figure}

We derived Fe abundances from the measured equivalent widths of 
$\sim$20-30 (for RGB stars) and $\sim$7-10 (for HB stars) unblended neutral lines 
using the code {\tt GALA} \citep{m13g}. 

Na abundances have been obtained 
by fitting the observed lines with a grid of synthetic spectra, in order to 
take into account the damped wings of the available Na lines. The [Na/Fe] abundances 
have been corrected for departures from local thermodynamical equilibrium
according to \citet{lind11}.
For the sake of consistency we applied the same corrections also to the Na abundances derived in 
\citet{mucciarelli16}.
Since the [Na/Fe] abundances of the two datasets are based on 
different Na lines some differences in the resulting abundances might be expected.
By using stars in common between the two data-sets we derived an average shift of ($+0.08\pm0.01$)dex,
which was added before combining all the available measures.

Uncertainties in the abundance ratios have been computed according to 
the procedure described in \citet{mucciarelli13}, which includes both the errors related to 
the measure of EWs and those arising from the atmospheric parameters.



\section{Sample selection}

Since our main goal is to compare the kinematic properties of FG and SG stars we adopted rather strict selection criteria 
to avoid contamination from spurious signals of any origin. 
For this reason we excluded from the final sample all the stars {\it (a)} lacking reliable Na abundance estimates (mainly due to low SNR 
or defects in the spectra), {\it (b)} with $\epsilon_{RV}$ anomalously large for their magnitude, 
namely $\epsilon_{RV}>0.8$~km s$^{-1}$ for V$\le 14.9$, $\epsilon_{RV}>1.8$~km s$^{-1}$ for $14.9<$V$\le 16.8$, 
and  $\epsilon_{RV}>2.8$~km s$^{-1}$ for V$> 16.8$, and {\it (c)} having $\sigma_{RV}>3.0$~km s$^{-1}$, 
as they are likely binary stars with large velocity amplitude. 

Figure~1 shows the distribution of RVs as a function of [Fe/H] and as a function of the projected distance from the cluster center (R) for the 632 stars 
that survived the above selection.
Stars belonging to NGC~6362 are easily identified in the $RV$-[Fe/H] diagram as they stand out from the field population 
at $\sim-15$ km s$^{-1}$ and [Fe/H]$\sim-1.1$. 
We selected as cluster members the stars within the dark grey box shown in Figure~1 (upper panel), 
enclosing stars with $-29.3$~km s$^{-1}$$\le RV\le+0.7$ km s$^{-1}$ and $-1.35<$[Fe/H]$<-0.8$. 
Finally, studying the velocity distribution of member stars as a function of R, we identified four stars whose 
$RV$ deviates from the systemic velocity of the cluster by more than three times the local value of the velocity dispersion, 
i.e. the velocity dispersion in a small radial range about their position.
Using these selection criteria, we selected a total of 489 bona fide member stars, 
which we define here as the {\it total sample}, plotted as black dots in Figure~1. 
Applying the maximum likelihood (ML) algorithm described in \citet{mucciarelli12} to the set of 489 individual [Fe/H] estimate 
and errors we find $\langle [Fe/H]\rangle =-1.063 \pm 0.004$ and $\sigma_{[Fe/H]}=0.000\pm 0.006$, in agreement with \citet{mucciarelli16}. 
It is reassuring that the outcome of our selection is a sample drawn from a pure single-metallicity population, 
thus very likely composed only by genuine cluster members.

The Na abundance distribution of member stars (Figure~2 panel a)) is clearly bimodal. This is in agreement  
with \citet{mucciarelli16} based on the much smaller sample (160 member stars) from the first observing run.
Using the Gaussian mixture modeling algorithm described by \citet{muratov10} we find that the hypothesis of unimodal distribution can be rejected with a probability $>99.9\%$. 
Based on the shape of the distribution shown in Figure~2 (panel a)) we define two sub-populations.
The first, here defined as {\it Na-poor sample}, includes stars with [Na/Fe]$<+0.05$ and numbers a total of 288 stars. The second,
{\it Na-rich sample}, is composed of stars
with [Na/Fe]$>+0.05$ for a total of 201 stars. 
We have checked that the adoption of slightly 
different boundaries 
does not affect our results (see Section~4). We verified that the two sub-samples have indistinguishable 
radial distributions consistent with their parent populations.
Based on our present understanding of MP formation, 
in the following we will sometimes refer to {\it Na-poor} and {\it Na-rich} sub-samples as FG and SG stars respectively.  

Panel b) of Figure~2 shows the distribution of Na-poor and Na-rich stars in the ($U, U-B$) CMD
\citep{dalessandro14}.  The two groups appear to be nicely separated on the RGB and in particular, 
Na-poor stars reside on the bluer side of the RGB (as they are N-poor),
while the Na-rich component lies on the red side, as expected \citep{sbordone11}.

\begin{figure}
\centering
\plotone{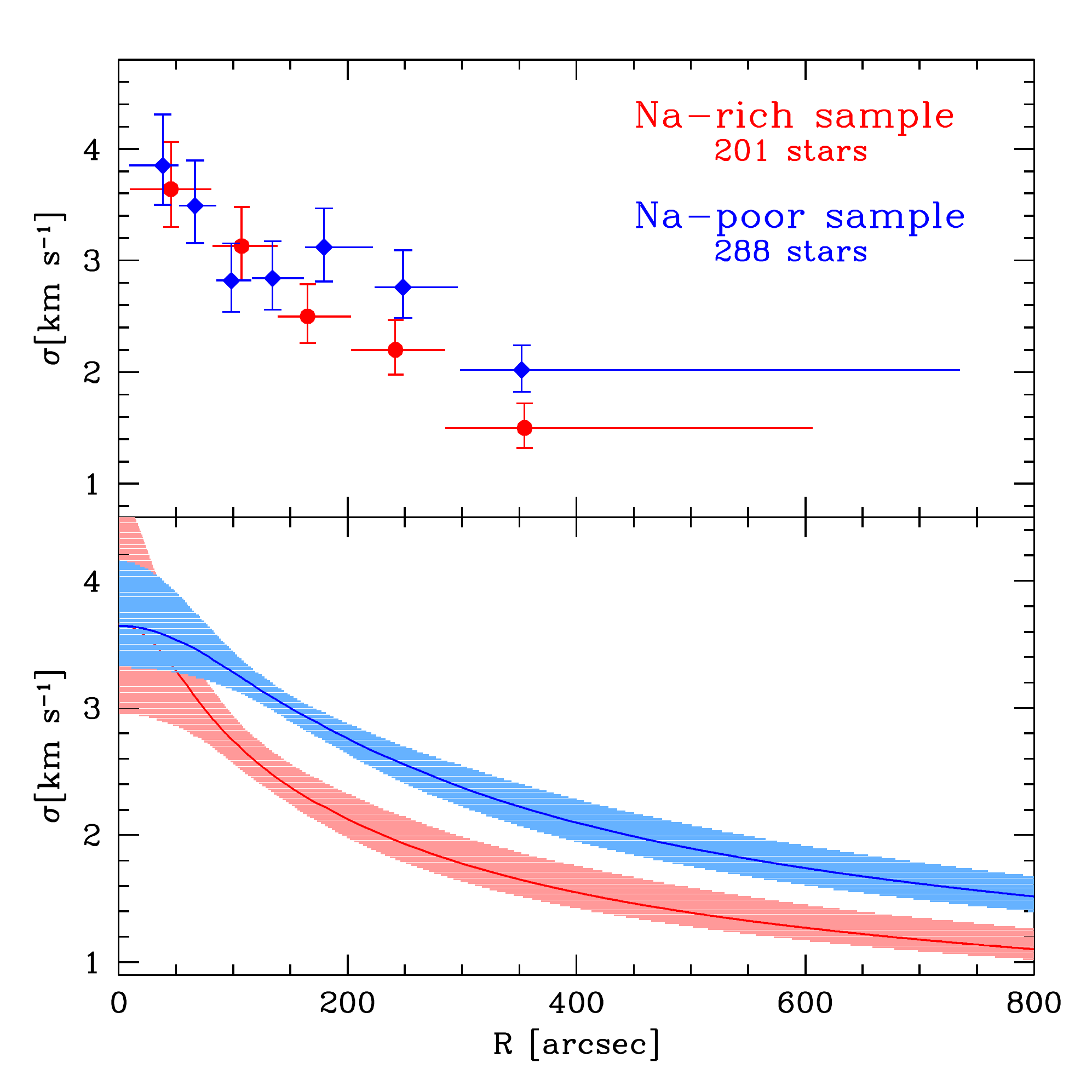}
\caption{Binned velocity dispersion profiles (upper panel) 
and 1-$\sigma$ confidence regions around the best-fit dispersion profiles (bottom panel) for
the {\it Na-rich} (red) and {\it Na-poor} (blue) samples.}
\label{sigma_profiles}
\end{figure}

\section{Kinematic analysis and Results}

In Figure~3 the line-of-sight velocity dispersion profiles of the two considered sub-samples are compared.
In the upper panel we show the binned dispersion profiles for illustration purposes only. 
They are obtained by assuming an equal number 
of stars  in each bin (apart from the outermost bin which contains any additional leftover stars). We used the 
Maximum Likelihood estimator of \citet{PM1993} to compute the velocity dispersion in each bin and its uncertainty. 
At each bin we assigned the value of the distance corresponding to the mean radius of all the stars in that bin. 
The corresponding horizontal error bars 
represent the radial range spanned by the stars in a given bin. 

To characterize the kinematics of the {\it Na-rich} and {\it Na-poor} samples, we used the maximum likelihood method described
in \citet{CorderoVHB2017}. 
We stress that this is a purely kinematic approach aimed at searching for relative differences in the kinematics of sub-populations and not aimed at providing a self-consistent dynamical model of the system.
For each sub-sample, we fit a kinematic model 
to discrete radial velocities. As in \citet{CorderoVHB2017}, we assume for the velocity dispersion profile the functional 
form of the \citet{Plummer1911} model, defined by its central velocity dispersion $\sigma_0$ and its scale radius $a$:

\begin{equation}\label{sigma2R}
\sigma^2(R) = \frac{\sigma_0^2}{\sqrt{1 + R^2/a^2}} \ ,
\end{equation}

where $R$ is the projected distance from the centre of the cluster. 
Table \ref{bestfit} lists the best-fit values and the uncertainties of the main parameters of our analysis 
and for the different samples considered.

\renewcommand{\arraystretch}{1.6} 
\begin{table*}
\caption{Median value and $\pm1 \, \sigma$ uncertainties (enclosing the central 68\% of the probability distribution) obtained from the posterior probability 
distributions for the free parameters of the different subsamples.}
\label{bestfit}
\begin{tabular}{l  c c c c}
\hline
Sample & $N_{\rm stars}$ & $v_{0}$ & $\sigma_0$ & $a$  \\
  & & [km~s$^{-1}$] &  [km~s$^{-1}$] & [arcmin]  \\
\hline
Na-poor & 288 & $-14.10^{+0.20}_{-0.19}$ & $3.55^{+0.38}_{-0.28}$ & $2.98^{+1.43}_{-0.96}$  \\
Na-rich & 201 & $-14.61^{+0.22}_{-0.22}$ & $3.61^{+0.60}_{-0.48}$& $1.69^{+1.11}_{-0.63}$   \\
Total sample& 489 & $-14.33^{+0.14}_{-0.14}$ & $3.61^{+0.32}_{-0.28}$ & $2.13^{+0.76}_{-0.55}$  \\

\hline
\end{tabular}
\end{table*}

In the bottom panel of Figure~3 the best-fit Plummer models and the 1-$\sigma$ confidence envelopes on the dispersion profiles are also shown. 

A comparison of the velocity dispersion profiles, in particular of those obtained without binning the data clearly shows differences between the velocity dispersion profile of the FG and the SG populations. In particular, while the two profiles are indistinguishable out to about $\sim 70\arcsec$-$80\arcsec$  (corresponding to $\sim$0.5$\times r_h$ and $\sim1.5 r_c$; \citealt{dalessandro14}), beyond this radius the SG velocity dispersion profile decreases more sharply than that of the FG and attains values of the dispersion smaller than those of the FG population by $\sim 1$ km s$^{-1}$.
This difference represents a large fraction ($\sim30\%$) of the observed 
central velocity dispersion values of {\it Na-poor} and {\it Na-rich} stars ($\sigma_0\sim3.55^{+0.38}_{-0.28}$ km s$^{-1}$ and $\sigma_0\sim3.61^{+0.60}_{-0.48}$ km s$^{-1}$ respectively)\footnote{On the other hand, we found no significant difference in the overall rotation pattern of the two sub-samples, adopting  
both the approach of \citet{Bellazzini2012} and that of \citet{CorderoVHB2017}. 
A more detailed analysis of the rotation properties of the cluster will be presented in a companion paper (E. Dalessandro et al. 2018, in preparation).} 
\footnote{We point out that we use a Plummer model just as a convenient way of providing a quantitative characterization of 
kinematical differences; the differences in this kinematical characterization are not meant to have any implication for 
the spatial distribution of the two populations which as shown in \citet{dalessandro14} are spatially mixed.} 

{\it We note that this is the first time that differences in the line-of-sight velocity dispersion of MPs are detected}.

In order to quantitatively assess the significance of this result, we 
performed a Kolmogorov-Smirnov test on the $RV$s. 
We find that the probability that the two populations 
are extracted from the same parent distribution is $P_{\rm KS}\sim4\times10^{-3}$.
\footnote{This test is not directly addressing 
the significance of the differences between the velocity dispersion profiles, but of the RV distribution.
In this respect it important to note that the shape of the cumulative distributions 
as well as the derived probabilities depend also on the
mean velocities of the two samples.}
We have also performed a {\it F-test} to test the hypothesis that the 
velocity dispersions of the two samples are equal. We used a maximum-likelihood analysis to derive 
the intrinsic velocity dispersions for the two sub-samples at different radial distances from the cluster center.  
We find that for R$>150\arcsec$ (which corresponds to the distance where velocity dispersion profiles Na-poor and Na-rich start 
to differ) the F-test probability is only $\sim4\times10^{-3}$ and it further decreases for R$>200\arcsec$ ($\sim4\times10^{-3}$).
{\it These results therefore provide strong support to the conclusion that the {\it Na-poor} and {\it Na-rich} velocity dispersion profiles  
are significantly different, with Na-poor stars being dynamically hotter.}

Considering the results of our previous study of MPs in NGC~6362 \citep{dalessandro14} showing that the spatial 
distributions of the FG and SG populations are consistent with being completely mixed, the kinematic difference we have found is surprising. Complete spatial mixing is expected in the advanced stages of a cluster's evolution and after the cluster has undergone significant mass loss. As we will further discuss in Section~6, once a cluster has attained complete spatial mixing, its MPs should also be characterized by similar velocity dispersion profiles. The results found here therefore raises a fundamental question concerning the dynamical ingredients responsible for the observed kinematic differences.

\section{FG and SG binary fractions}

Before exploring the possible culprit of the observed kinematic differences, we use our radial velocity data to determine the binary fraction in the FG and SG populations. Information about the binary population and the possible differences between the FG and SG binary fractions is extremely important both to build a complete dynamical 
picture of the cluster and to shed light on the possible role of binaries on the observed kinematic properties presented in the previous section.  The finding by \citet{lucatello15} that FG stars have typically a larger binary fraction than their SG counterparts seem particularly relevant in this context.

We used a sub-sample of 384 stars observed repeatedly (from a minimum of 
2 to a maximum of 6 times; see Section~2) within a period of 704.96 days ($\sim$2 yrs). Candidate binary stars 
(i.e. those with $\sigma_{RV}>3$ km s$^{-1}$ -- see Section~3) have now been included in the analysis.
We stress that this is the largest sample ever used so far for such kind of study for an individual GC.

We followed the approach described in \citet{lucatello15} for the binary fraction derivation. 
For each star in our sub-sample, we performed a $\chi^{2}$ test 
by using the single velocities and relative errors 
to assess whether they are compatible with a non-variable 
behavior. Stars with $P(\chi^{2})<1\%$ have been flagged as binaries.
Among the 384 stars (235 
belonging to the FG and 149 to the SG), 12 
turned out to be binaries, 11 are FG stars and 1 belong to the SG sub-population. These values correspond to a minimum 
binary fraction for the entire sample (FG+SG) $f_{min}^{TOT}=3.1\pm0.9\%$, and to $f_{min}^{FG}=4.7\pm1.4\%$ and 
$f_{min}^{SG}=0.7\pm0.7\%$.   

Unfortunately, a significant fraction of binaries cannot be directly detected with our data-set
because of their long variability periods and/or not favorable 
inclination angles.
In order to account for these observational limitations and estimate the binary detection efficiency of 
our observations, we adopted the following approach.
For each observed star, we generated a synthetic population of 1000 binaries 
assuming a primary component mass of 0.85 $M_{\odot}$ (adequate for a 
RGB star in NGC 6362), secondary component masses randomly extracted 
from a flat mass-ratio distribution, periods extracted 
from a log-normal distribution (using $\langle \log (P/d)\rangle=4.8$ and 
$\sigma_{\log (P/d)}=2.3$; \citealt{DM91}), eccentricities extracted 
using the prescriptions of \citet{DM91}, random inclination angles, periastron longitude 
and orbital phases. We also forced the binary semi-major axes 
to lie between the Roche-Lobe overflow distance limit \citep{lee88} and that 
corresponding to the average collisional ionization limit \citep{hut1983}. 
The adoption of this additional criterion reduces the extent of the simulated period distribution to the range 
$0.4<\log (P/d)<5.3$.
For each synthetic binary, the luminosity weighted systemic velocities has been computed and it
has been sampled with the same cadence as the observations.

Gaussian shifts with standard deviation equal to observational errors have been 
added to the velocities to mimic the effect of observational uncertainties. 
As a control population, we also simulated a large number of single stars with 
constant velocity.

The same analysis performed on the observed sample has been applied to 
the synthetic populations of binaries and single stars to derive the 
detection efficiency as a function of period and the false detection frequency.
The detection efficiency ranges from $\sim40\%$ at log(P/d)=0.6 to $\sim60\%$ at 
log(P/d)=1.6 with a decreasing tail at large periods reaching zero at log(P/d)$>4.5$.
Overall, an average detection 
efficiency of $\sim26\%$ has been found, while the false detection frequency is
$\sim1\%$. FG stars have a slightly larger detection efficiency with respect to SG ones 
(26.7\% versus 19.5\%), likely due to the smaller velocity measurements errors 
in FG stars characterized by strong Na lines. Based on these results, the global binary fractions of FG 
and SG turn out to be $f_{g}^{FG}\sim14.3\%$ and $f_{g}^{SG}<1\%$ respectively.

This analysis, which is based on a sample up to 8 times larger (per individual cluster), confirms
previous findings \citep{lucatello15} about the different present-day binary fraction of FG and SG sub-populations. It also provides support to the predictions of the theoretical studies of \citet{vesperini11} and \citet{hong15,hong16} showing that the SG binary fraction is expected to be smaller than that of the FG population as a result of dynamical evolution.

\subsection{The effect of binary fraction differences on the velocity dispersion profiles}
To estimate the potential effect of such different binary fractions on the derived 
velocity dispersion profiles of FG and SG sub-populations, we first adopted the following approach. 
Mock observations have been constructed 
by randomly extracting for each observed star a velocity from a gaussian function 
with dispersion equal to the line-of-sight velocity dispersion profile at the same cluster-centric distance of the observed star.  The same line-of-sight 
velocity dispersion profile has been adopted for FG and SG stars, following the 
\citet{king1966} model that best fits the cluster density profile (see \citealt{dalessandro14}).
A random number ($rn$) comprised in the range 0-1, extracted from a uniform probability distribution, 
was assigned to each FG star. 
For stars with $rn<0.143$, which corresponds to the estimated global FG binary fraction, 
a synthetic binary has been simulated with the mass ratios and orbital parameters described above and 
the corresponding mean velocity shift has been added. 
The LOS velocity dispersions of the two generations have been 
then computed using {\it a)} the entire sample and {\it b)} only stars with R$>150\arcsec$ (i.e. where the maximum difference has been observed).
The same selection criteria adopted for the construction of the observed dispersion profiles 
and to exclude apparent binaries has been applied in the 
analysis of the mock observations. A set of 1000 extractions 
has been performed for both the considered radial ranges and the distribution 
of differences has been derived. The results are shown in Figure~4. It can be seen that a 
velocity dispersion difference equal or larger than the observed one can be 
obtained as a result of the effect of the different binary fractions in 
$\sim54\%$ of the simulated cases (if the entire radial range is considered) 
and in $\sim28\%$ of the cases when only the outermost radial range is considered.
{\it Hence, the observed difference in the velocity dispersion profile between FG ad SG stars in NGC~6362 is consistent} 
with being due to the difference in the binary fraction of the two populations.

\begin{figure}
\centering
\plotone{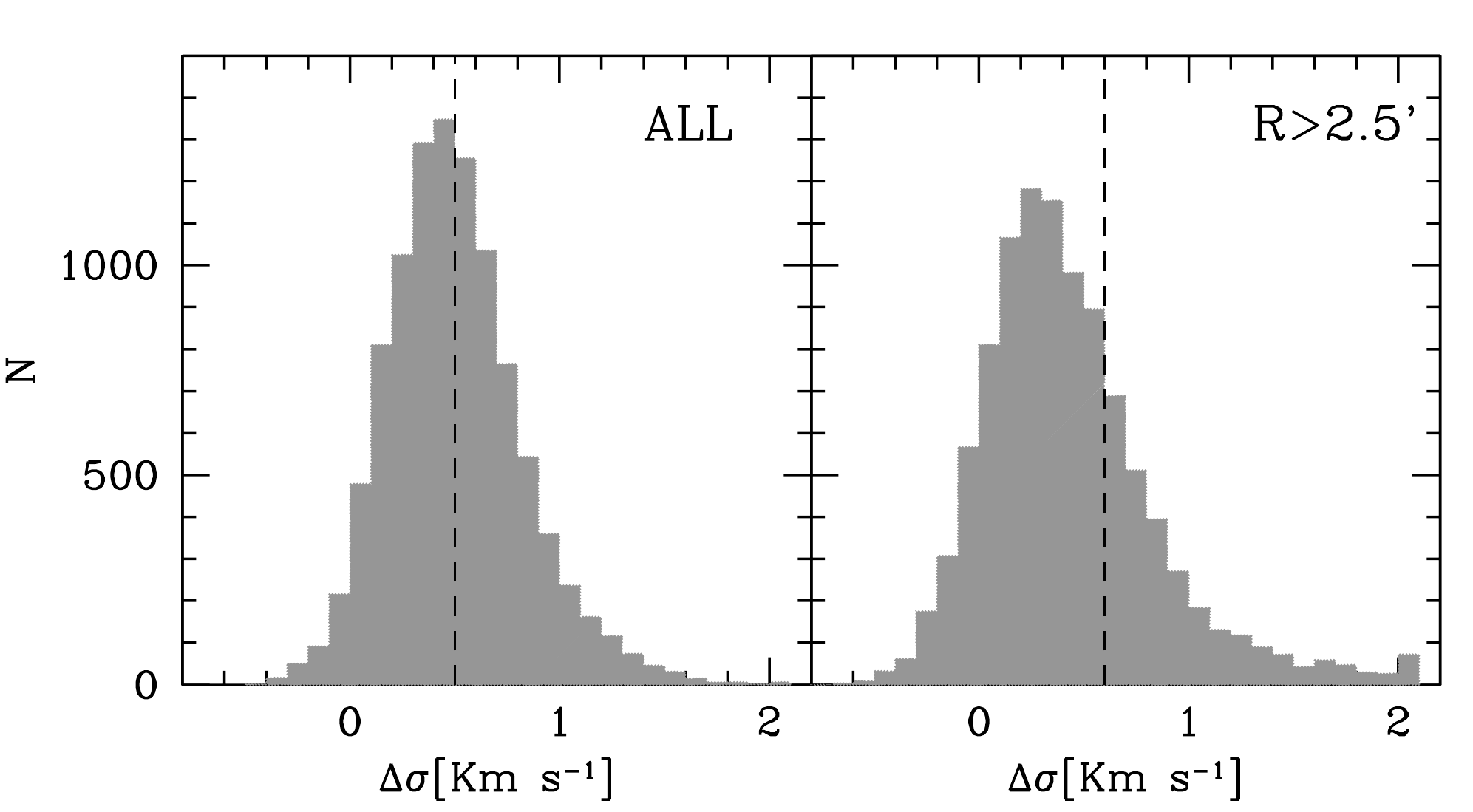}
\caption{Distributions of the velocity dispersion differences obtained as described in Section 4.1 for the entire
sample and for stars located at $R>2.5\arcmin$ from the cluster center. Dashed lines represent the observed velocity dispersion differences.}
\label{sigma_profiles}
\end{figure}

\begin{figure}
\centering
\plotone{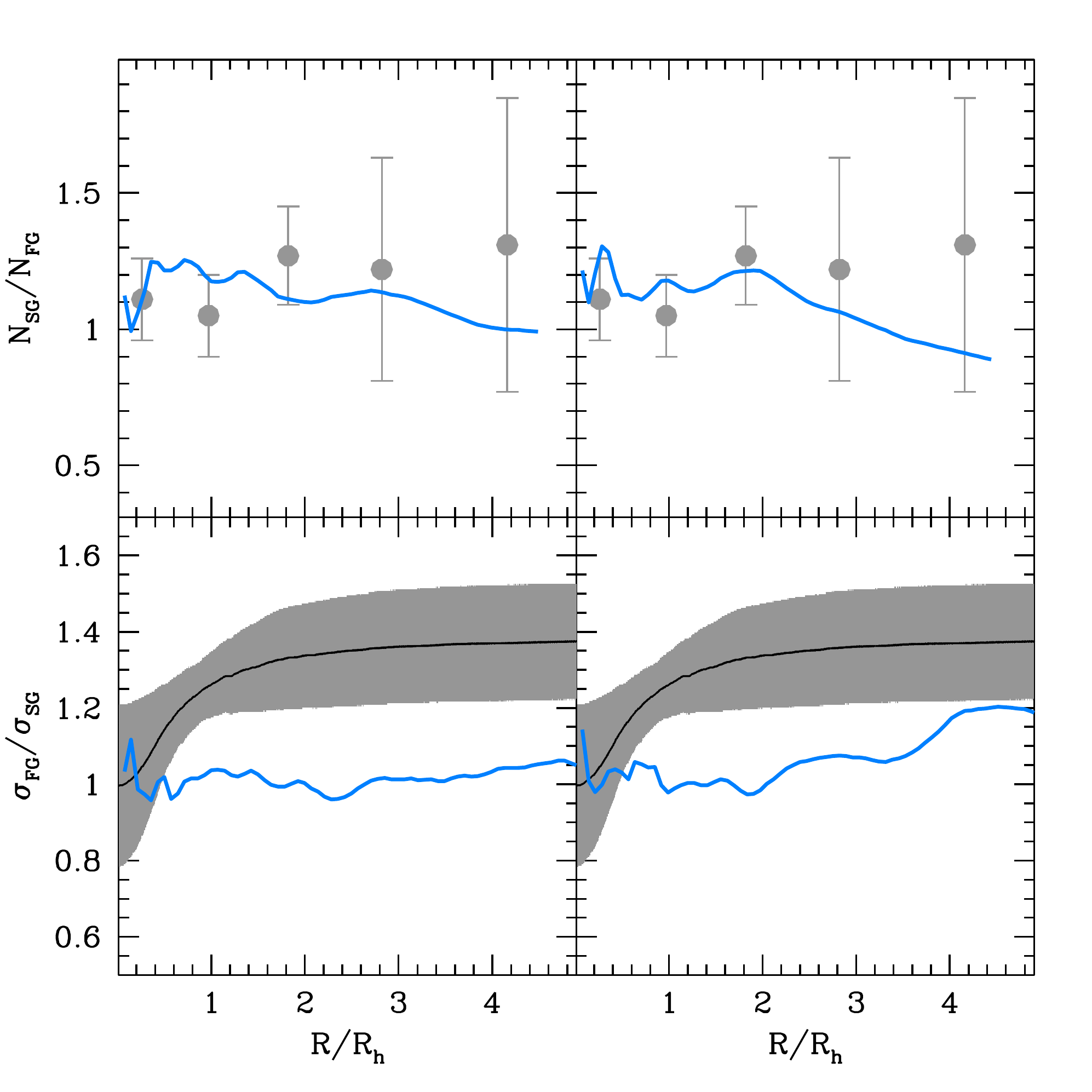}
\caption{{\it Upper panels}: FG and SG number ratio as a function of the distance from the cluster center normalized to $R_h$ in two advanced snapshots of our $N$-body simulations with no binaries (blue lines). The grey circles represent the number ratio distribution observed in NGC~6362 \citet{dalessandro14}. 
{\it Lower panels}: Radial variations of the ratio of the observed FG to SG line-of-sight velocity dispersions ($\sigma_{\rm FG}/\sigma_{\rm SG}$) from $N$-body models for the same snapshots as before. The dashed grey area represents the observations.}
\label{sigma_profiles}
\end{figure}

\section{$N$-Body models}

In order to further explore the possible dynamical history behind the kinematical  differences revealed by our observations, we have studied the evolution of the line-of-sight velocity dispersion of FG and SG stars in an $N$-body simulation.
The goal here is not to present a model specifically tailored to fit in detail the dynamics of NGC~6362, but rather to gather some fundamental insight on the dynamical ingredients necessary to explain the observational results.
For details on the initial conditions of the simulation we refer to \citet{vesperini18}, here we just summarize the main points. Our simulation starts with a SG population embedded within a more extended FG population: the initial half-mass radius of the FG population is about five times larger than that of the SG population. The simulation starts with 50,000 stars equally split between FG and SG. In this simulation we have focussed our attention on the effects of two-body relaxation on 
the evolution of the spatial and kinematical properties of the two populations.

In Figure~5 we show the radial profile of the number ratio of SG to FG stars (upper panels) and the ratio profiles of the FG to SG line-of-sight velocity dispersions ($\sigma_{\rm FG}/\sigma_{\rm SG}$; lower panel) representative of two advanced evolutionary stages in our simulation for stars with masses between $0.75-0.85 M_{\odot}$.  
Although as pointed out above, our analysis is not aimed at providing a detailed fit of the observations, we include in these plots also the observational data from the analysis carried out in this paper and in \citet{dalessandro14} to ease  a more quantitative comparison between the strength of the observed and theoretical gradients.

The plots of Figure~5 shed light on the close connection between the structural and kinematical properties of the FG and SG populations. 
For a dynamically old cluster in which the FG and the SG populations are completely spatially mixed (Figure~5 left panels), 
the $N$-body simulations show that the radial profile of $\sigma_{\rm FG}/\sigma_{\rm SG}$ is flat, at odds with the observations. 
If we consider a dynamical phase characterized by a modest radial gradient in the fraction of SG stars 
(but still consistent within the errors with the observations; Figure~5 right panels), 
the $\sigma_{\rm FG}/\sigma_{\rm SG}$ tends to increase in the external regions of the cluster, 
but the values of $\sigma_{\rm FG}/\sigma_{\rm SG}$ remain smaller than the observed ones at all distances from the cluster center.  
 
Although the discrepancy between the theoretical and the observed radial profile of $\sigma_{\rm FG}/\sigma_{\rm SG}$ can be of the order of $\sim1.5\sigma$ in the external regions, the systematic underestimate of the ratio compared to the observed one could be an indication that an additional dynamical effect not accounted for in the simulations might be responsible for the observed profile. Indeed, as discussed in Section~5.1, a difference between the FG and SG binary fractions can play a key role in determining the observed differences between the FG and SG velocity dispersion profiles.

\begin{figure}
\plotone{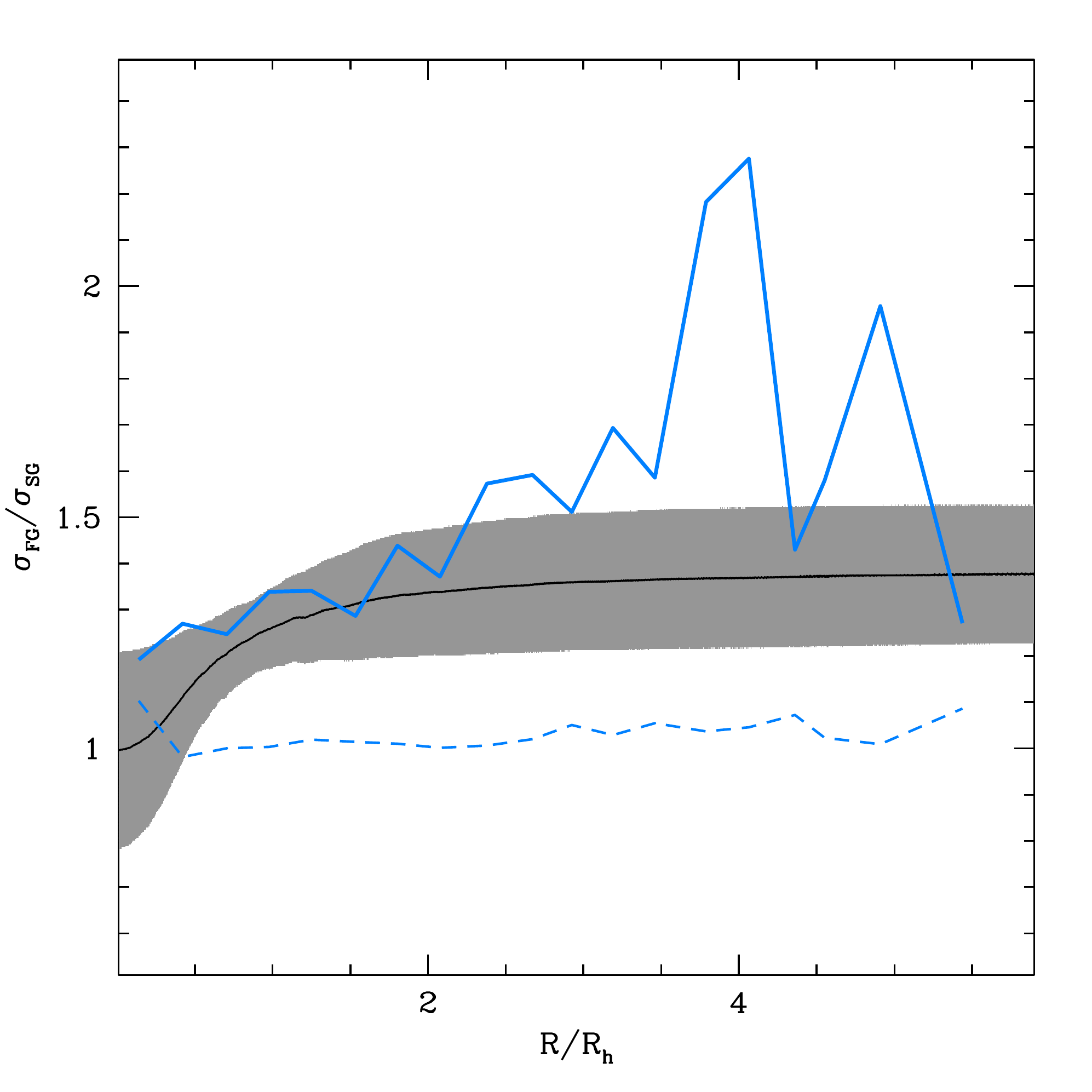}
\caption{Radial variations of the ratio of the observed FG to SG line-of-sight velocity dispersions ($\sigma_{\rm FG}/\sigma_{\rm SG}$) 
from $N$-body models with binaries (see Section~6) compared to observations (grey dashed area). The dashed line
represents the velocity dispersion ratio radial distribution from the same simulation when the effect of binaries is not included. }
\end{figure}

As shown in \citet{vesperini11} and \citet{hong15,hong16}, if SG stars form in a more compact and centrally concentrated subsystem than FG stars, as predicted by a number of formation models (see e.g. \citealt{dercole08}), all the processes altering the number and orbital properties of binary stars (ionization, hardening, softening, ejection; see e.g. \citealt{heggie_hut03}) affect the SG binaries more efficiently than the FG ones. One of the consequences of this dynamical difference is the preferential disruption and ejection of SG binaries leading, in turn, to a larger global fraction of FG binaries.

As shown in the study of \citet{hong16} and Hong et al. (2018, in preparation), the processes affecting the evolution and survival of binaries have also an effect on their spatial distribution and the spatial mixing of the FG and SG binaries. 
In particular, the timescale for the spatial mixing of FG and SG binaries 
can be much longer than that of single stars. This implies that while the FG and SG single stars might have already reached complete spatial mixing, the FG and SG binary populations might still be characterized by a radial gradient, with the fraction of FG to SG  binaries increasing as the cluster-centric radius increases (see e.g. bottom panel of Figure~11 in \citealt{hong16}). The difference in the fraction of FG and SG binaries and its radial variation imply that the possible velocity dispersion inflation due to binaries is stronger for the FG population and is increasingly more important at larger distances from the cluster center. As a consequence, differences in the FG and SG binary fraction and in their spatial distribution can contribute to produce a $\sigma_{\rm FG}/\sigma_{\rm SG}$ profile increasing with the distance from the cluster center, as found in our observations. 

Figure~6 illustrates this effect as measured in one of the simulations (MPr5f1x3-800) presented in \citet{hong16}. We emphasize that the simulations are still idealized and are not meant to provide a detailed model for NGC~6362 but they include the essential dynamical ingredients necessary to illustrate the effect 
of interest here. 
From this analysis and the comparison with the observations it emerges that binaries 
can play a major role in shaping the $\sigma_{\rm FG}/\sigma_{\rm SG}$ radial gradient and could be the dynamical
ingredient needed to match the observational results.

Although additional and more realistic simulations are needed to build specific models for NGC~6362, the results presented here clearly illustrate how the study of the kinematics of multiple populations can reveal the fingerprints of a number of fundamental dynamical effects and of their role in shaping the properties of FG and SG stars.

\section{Summary and Discussion}

The detailed kinematic analysis performed in this work has revealed that Na-poor (FG) and Na-rich (SG) stars 
are characterized by significantly different line-of-sight velocity dispersion profiles. SG stars have systematically smaller velocity 
dispersion values than FG ones, with differences of $\sim 1$ km s$^{-1}$  for $R>70\arcsec-80\arcsec$ (corresponding to $R>0.5 r_{\rm h}$). 
This is the first time that differences in the line-of-sight velocity dispersion of MPs are detected.

Considering that in our previous study on the spatial distribution of MPs in NGC~6362 \citep{dalessandro14} we have found that the FG and SG populations are spatially mixed and that the cluster must be in an advanced stage of its dynamical evolution, the kinematical evidence detected here is surprising and raises a fundamental question concerning the dynamical processes responsible for the difference between the FG and SG velocity dispersion profiles.

Thanks to our large set of RVs we have also been able to estimate the binary fraction in the two populations and found a significant difference between the FG binary fraction ($f\sim14\%$) and that of the SG population ($<1\%$).
This result is based on the largest sample ever used for this kind of analysis. 
 
By using $N$-body simulations and mock observations, we show that such a large binary fraction difference 
can play an essential role in determining the observed kinematic differences between the FG and the SG populations found in our study.

Beside the specific case of NGC~6362, the results of this paper clearly demonstrate the importance 
of the study of the kinematics at several epochs to build a complete dynamical picture of MPs in globular clusters and to shed 
light on the dynamical history of MPs.
In this context it will be important to extend this kind of analysis to other systems in order to understand whether NGC~6362 is a peculiar case or similar effects are present in all GCs. Moreover, the addition of Gaia proper motions sampling the entire extension of the cluster,
will allow us to constrain the degree of anisotropy currently characterizing the system.

More in general, for clusters at different dynamical stages a radial variation of the SG to FG velocity dispersion could be due to a combination of the effect of binaries and a radial gradient in the fraction of SG stars.

Moving a step forward in our comprehension of the kinematics of MPs is, in turn, a key stage in the study of GC formation and evolution.
 
\vspace{1.5cm}
The authors thank the referee Mario Mateo for the careful reading of the paper and useful comments that improved the
presentation of this work.
ED aknowledges support from The Leverhulme Trust Visiting Professorship Programme VP2-2017-030.



\end{document}